\documentclass[allclo]{FBSart}

\def\bea{\begin{eqnarray}}
\def\eea{\end{eqnarray}}
\def\be{\begin{equation}}
\def\ee{\end{equation}}

\def\bra{\langle}
\def\ket{\rangle}

\title{Nonperturbative solution of Yukawa theory and gauge theories%
\footnote{Presented at Light Cone 2004, Amsterdam, 16 - 20 August.}}

\author{J.R. Hiller}
\institute{Department of Physics, University of Minnesota-Duluth, Duluth, MN 55812 USA}
\runningauthor{J.R.\,Hiller}
\runningtitle{LC 2004}
\begin{document}
\maketitle
\begin{abstract}
Recent progress in the nonperturbative solution of 
(3+1)-dimensional Yukawa theory and quantum electrodynamics
and (1+1)-dimensional super Yang--Mills theory is summarized.
\end{abstract}
\section{Introduction}

Field theories can be solved nonperturbatively when quantized on
the light cone~\cite{Dirac,PauliBrodsky,DLCQreview}.  This is done
in a Hamiltonian formulation which, unlike Euclidean lattice
gauge theory~\cite{lattice}, yields wave functions.
The properties of an eigenstate can then be 
computed relatively easily.  Success came easily for
two-dimensional theories~\cite{DLCQreview}, but in three or four
dimensions the added difficulty of regularization and renormalization
has until recently limited the success of the approach.
Here we discuss recent progress with two different yet related
approaches to regularization.  One is the use of Pauli--Villars (PV)
regularization~\cite{bhm-previous} and the 
other, supersymmetry~\cite{SDLCQreview}.

\section{Pauli--Villars regularization}

The most important aspect of the PV approach is the introduction of
negative metric PV fields to the Lagrangian, with couplings
only to null combinations of PV and physical fields.
This choice eliminates instantaneous
fermion terms from the Hamiltonian and, in the case of QED,
permits the use of Feynman gauge without inversion of a covariant
derivative.  
In addition, the Hamiltonian eigenvalue problem can be formulated
in terms of transverse polar coordinates, which allow 
direct construction of
eigenstates of $J_z$ and explicit factorization from the
wave function of the dependence on the polar angle; this
reduces the effective space dimension and the size of the
numerical calculation.  The numerical approximation
requires the introduction of special discretizations
rather than the traditional
momentum grid with equal spacings used in discrete light-cone
quantization (DLCQ)~\cite{PauliBrodsky,DLCQreview}.  
The special discretization allows
the capture of rapidly varying integrands in the product
of the Hamiltonian and the wave function, which occur for
large PV masses.

An analysis of Yukawa theory with a PV scalar and a PV fermion,
but without antifermion terms, is given in Refs.~\cite{OneBoson}
and \cite{bhm}.
No instantaneous fermion terms appear in the light-cone
Hamiltonian, because they are
individually independent of the fermion mass and cancel
between instantaneous physical and PV fermions.
For a positive-helicity dressed fermion state $\Phi_+$,
the wave functions satisfy the coupled 
system of equations that results from the Hamiltonian eigenvalue
problem $P^+P^-\Phi_+=M^2\Phi_+$.  Each wave function has a total 
$L_z$ eigenvalue of 0 (1) for a constituent-fermion helicity
of $+1/2$ ($-1/2$).

Truncation to one boson leads to an analytically solvable 
problem~\cite{OneBoson}.  The one-boson--one-fermion wave 
functions then have simple explicit forms, and
the bare-fermion amplitudes must obey an
algebraic equation which can be readily solved.
An analysis of this solution is given in Ref.~\cite{OneBoson}.

In a truncation to two bosons, we obtain reduced
equations for the one-boson--one-fermion wave functions~\cite{bhm}.
These reduced integral equations are converted to a matrix equation 
via quadrature and then diagonalized.  The diagonalization 
yields the bare coupling as an eigenvalue and 
the discrete wave functions from the eigenvector.
From the wave functions we can obtain any property of the state.

We apply these same techniques to QED in Feynman gauge~\cite{qed}.
In addition to the absence of instantaneous fermion interactions,
we find that the constraint equation for the nondynamical fermion
field is independent of the 
gauge field and can therefore be solved without inverting
a covariant derivative.  The resulting Hamiltonian is given in
Ref.~\cite{qed}.

We consider the dressed electron state, without pair contributions and 
truncated to one photon.  The one-photon--one-electron wave functions 
again have simple forms, and the bare-electron amplitudes satisfy
algebraic equations nearly identical to those found in Yukawa theory.  
An analytic solution is again obtained.  From this solution we can
compute various quantities, including the anomalous magnetic 
moment~\cite{qed}.

\section{Supersymmetric theories}

For super Yang--Mills (SYM) theory, the technique used is supersymmetric
discrete light-cone quantization (SDLCQ)~\cite{Sakai,SDLCQreview}. 
This method is applicable to theories with enough supersymmetry to
be finite.  The supersymmetry is maintained exactly within the 
numerical approximation by discretizing
the supercharge $Q^-$ and computing the discrete Hamiltonian
$P^-$ from the superalgebra anticommutator $\{Q^-,Q^-\}=2\sqrt{2}P^-$.
To limit the size of the numerical calculation, we work in the
large-$N_c$ approximation; however, this is not a fundamental
limitation of the method.  

The stress-energy correlation function for ${\mathcal N}$=(8,8) SYM theory 
can be calculated on the string-theory side~\cite{N88correlator}:
$F(x^-,x^+)\equiv\bra T^{++}(x)T^{++}(0)\ket=(N_c^{3/2}/g)x^{-5}$. 
We find numerically that this is {\em almost} true in 
${\mathcal N}$=(2,2) SYM theory~\cite{N22}.
By analyzing the Fourier transform with respect to the total
momentum $P^+=K\pi/L$, where $K$ is the integer
resolution and $L$ the length scale of DLCQ~\cite{PauliBrodsky}, we 
obtain~\cite{N22}
\be  
   F(x^-,x^+)=\sum_i \Big|\frac L{\pi}\bra 0|T^{++}(K)|i\ket\Big|^2\left(
   \frac {x^+}{x^-}\right)^2 \frac{M_i^4}{8\pi^2K^3}K_4(M_i \sqrt{2x^+x^-}) .
   \label{cor}
\ee
We then continue to 
Euclidean space by taking $r=\sqrt{2x^+x^-}$ to be real. 
The matrix element $(L/\pi)\bra 0|T^{++}(K)|i\ket$ is independent of $L$.
Its form can be substituted directly to give an explicit expression for the 
correlator. 

The correlator behaves like $(1-1/K)/r^4$ at small $r$.
For arbitrary $r$, it can be obtained numerically by either 
computing the entire spectrum (for ``small'' matrices)
or using Lanczos iterations (for large)~\cite{correlator}.
We find~\cite{N22} that for intermediate values of $r$,
the correlator behaves like $r^{-4.75}$,
or almost $r^{-5}$. The size of this intermediate region increases
as $K$ is increased.

We next consider ${\mathcal N}$=(1,1) SYM theory at finite 
temperature~\cite{FiniteTemp}.
From the discrete form of the supercharge $Q^-$
we can compute the spectrum, which at large-$N_c$
represents a collection of noninteracting modes.  With a sum over
these modes, we can construct the free energy at finite temperature
from the partition function~\cite{Elser} $e^{-p_0/T}$.
We obtain the total free energy as~\cite{FiniteTemp} 
\be
{\mathcal F}(T,V)=-\frac{(K-1)\pi}{4}
   VT^2-\frac{2VT}{\pi}\sum_{n=1}^{\infty}
    {\sum_{l=0}^{\infty}}M_{n}\frac{K_{1}
      \left((2l+1)\frac{M_{n}}{T}\right)}{(2l+1)} .
\ee
The sum over $l$ is well approximated by the first few terms.
We can represent the sum over $n$ as an integral over a density 
of states: $\sum_n \rightarrow \int \rho(M) dM $
and approximate $\rho$ by a continuous function.
The integral over $M$ can then be computed by standard numerical 
techniques.  We obtain $\rho$ by a fit to the computed spectrum 
of the theory and find $\rho(M)\sim \exp(M/T_{\rm{H}})$,
with $T_H \sim 0.845\sqrt{\pi/g^2N_c}$, the Hagedorn 
temperature~\cite{Hagedorn}.  From the free energy we can compute 
various other thermodynamic functions up to this 
temperature~\cite{FiniteTemp}.

\section{Future work}

Given the success obtained to date, these techniques are well worth
continued exploration.  In Yukawa theory, we plan to consider the 
two-fermion sector, in order to study true bound states.  For QED
the next step will be inclusion of two-photon states in the calculation
of the anomalous moment.  For SYM theories, we are now able to reach
much higher resolutions; this will permit
continued reexamination of theories where previous calculations were 
hampered by low resolution.

\section*{Acknowledgments}
The work reported here was done in collaboration with 
S.J. Brodsky, G. McCartor, V.A. Franke, S.A. Paston, 
and E.V. Prokhvatilov, and
S. Pinsky, N. Salwen, M. Harada, and Y. Proestos,
and was supported in part by the US Department of Energy
and the Minnesota Supercomputing Institute.

\end{document}